\documentclass[
12pt,
]{article}
\usepackage{fullpage}

\begin{document}

\message{<< Assuming 8.5" x 11" paper >>}    

\raggedbottom

\parskip=9pt

\def\singlespace{\baselineskip=9pt}      
\def\sesquispace{\baselineskip=12pt}      


\def\alfa{\alpha}
\font\openface=msbm10 at10pt
\def\Minkowski     {{\hbox{\openface M}}}
\def\NaturalNumbers{{\hbox{\openface N}}}

\def\tilde{\widetilde}		

\font\titlefont=cmb10 scaled\magstep2 
\font\subheadfont=cmssi10 scaled\magstep1 
\def\author#1 {\medskip\centerline{\it #1}\bigskip}
\def\address#1{\centerline{\it #1}\smallskip}
\def\furtheraddress#1{\centerline{\it and}\smallskip\centerline{\it #1}\smallskip}
\def\email#1{\smallskip\centerline{\it address for email: #1}}
\def\eprint#1{{\tt #1}}
\def\reference{\hangindent=1pc\hangafter=1} 
\def\ref{\reference}
\def\AbstractBegins
{
 \singlespace                                        
 \bigskip\leftskip=3truecm\rightskip=3truecm     
 \centerline{\bf Abstract}
 \smallskip
 \noindent	
 } 
\def\AbstractEnds
{
 \bigskip\leftskip=0truecm\rightskip=0truecm       
 }

\def\ReferencesBegin
{
 \singlespace					   
 \vskip 0.5truein
 \centerline           {\bf References}
 \par\nobreak
 \medskip
 \noindent
 \parindent=2pt
 \parskip=6pt			
 }

\def\journaldata#1#2#3#4{{\it #1\/}\phantom{--}{\bf #2$\,$:} $\!$#3 (#4)}

\def\sepref{\baselineskip=18pt \hfil\break \baselineskip=12pt}

\def\subsection #1 {\medskip\noindent{\subheadfont #1
 }\par\nobreak\noindent}

\def\linebreak{\hfil\break}
\def\lbr{\linebreak}

\def\dash{ --- }

\def\versionnumber#1{
 \vskip -1 true in \medskip 
 \rightline{version #1} 
 \vskip 0.3 true in \bigskip \bigskip}



\phantom{}




\sesquispace

\centerline {\titlefont On Cosmological Constant in Causal Set Theory } 
\bigskip


\singlespace			        

\author{Yevgeniy Kuznetsov}

\email{ykuznetsov@physics.ucsd.edu}

\AbstractBegins                              

Resolution of the cosmological constant problem based on Causal Set theory is discussed. It is argued that one should not observe any spacetime variations in $\Lambda$ if Causal Set approach is correct.

\AbstractEnds

\sesquispace

\vskip 30pt

\noindent

\normalfont
It's been known since late 1990's that expansion of our Universe appears to be accelerating. [1] The most commonly agreed upon explanation of this expansion is a small but non-zero cosmological constant.

The cosmological constant, $\Lambda$, appears as an additional term in Einstein's field quations:

$$R_{\alpha\beta} - {1 \over 2} R g_{\alpha\beta} + \Lambda g_{\alpha\beta} = \kappa T_{\alpha\beta}$$

$$\kappa = 8 \pi G / c^4$$

It has dimensions of inverse length squared. Currently, the best estimate of its value is

$$\Lambda = {3 H^2\over c^2} \Omega_{\lambda} = (1.29 \pm 0.08)*10^{-52} m^{-2}$$

This value is extremely low ($\sim 10^{-120}$ in Planck units), and there does not seem to be any natural way to explain it without invoking "`new"' physics or metaphysical arguments such as antropic principle. One could hope to explain a cosmological constant that is \textit{exactly zero}, but this situation appears to be ruled out by experimental data.

An interesting explanation of small but non-zero $\Lambda$ was proposed by Rafael Sorkin [2]. It is based on the theory of Causal Sets. 

Causal Set theory is, in some sense, an "`envelope theory"' for discrete quantum gravity theories. Discrete quantum gravity paradigm is an approach to quantization of gravity that attempts to go around non-renormalizability of traditional path integral of quantum gravity by postulating that the spacetime is discrete, and one should sum over a countable number of unique discrete structures (or, at worst, perform a finite number of regular integrals) rather than performing a functional integration over (ill defined) space of possible metrics. "`Path integral"' (or, more properly, "`sum over histories"') is written as follows:

\begin{equation}
Z = \sum_{configurations} \exp{(iS/\hbar)} + \textrm{boundary terms}
\end{equation}

where $S$ is the discretized version of the Einstein-Hilbert action functional:

\begin{equation}
S = \int (R + \kappa L_{matter} - 2\Lambda) dV = \int (R + \kappa L_{matter}) dV - 2 \Lambda V
\end{equation}

Here $R$ is the curvature scalar, $L_{matter}$ is the matter Lagrangian, and $dV$ is the volume element:

$$dV = \sqrt{-g} d^4x$$

We perform summation over the space of all possible spacetime configurations and matter fields, modulo diffeomorphisms. 

Two results of Causal Sets are used in derivation of the expression for the cosmological constant: first, the assumption that the spacetime is intrinsically discrete, with volumes of its elements on the order of Planck volume; and, second, the assumption that total 4-volume of a set of $N$ elements approximately $Nl_{pl}^4$, and fluctuates depending on the exact structure of relationships between them. These fluctuations are Poisson-like, on the order of $\sqrt{N}$. 

The theory of gravity that is expected to emerge from Causal Sets in the continuum limit is not the usual Einstein's GR, but its reinterpretation called "`unimodular gravity"'. In unimodular gravity, we restrict our attention to constant 
volume configurations ($g = \det g_{\mu\nu} = -1$). If volume is constant, $\Lambda$ clearly drops out of the functional, having no effect on dynamics of the system.

The argument now goes as follows. In discrete limit gravity we should consider configurations with fixed number of elements $N$, rather than fixed volume. Therefore our sum-over-histories will receive contributions from a range of $V$. Since $V$ and $\Lambda$ are conjugate, uncertainties in them will be related:

$$\Delta V \Delta \Lambda \sim \hbar$$

i.e. 

\begin{equation}
|\Delta\Lambda| \sim 1/\sqrt{V}
\end{equation}

At the Hubble scale, the resulting fluctuation is on the order of $1/H^2$, which is in agreement with observations.

The basic result we obtain is quite simple. Its implementation and interpretation, however, is much more problematic, not the least because we don't yet have the complete quantum dynamics of causal sets. 

An important question with regards to this explanation of cosmological constant is, what does $\Lambda$ fluctuate with respect to? This question is typically answered very vaguely in existing literature. It is often suggested that $\Lambda$ would somehow gradually vary over time (i.e. the relevant volume is equal to the volume of the past directed light cone at all times). [3] In response, it's been argued in [4] that such variations would lead to considerable anisotropy in CMB radiation. Furthermore, nothing in the argument made above makes any references to light cones. An equally plausible interpretation could be that $\Lambda$ is to fluctuate as $1/\sqrt{V}$ when averaged over \textit{any} 4-volume (although this would lead to gross deviations from flatness of spacetime at laboratory scales: "`dark energy density"' fluctuations in a 1x1x1 m box would be on the order of $\rho_{crit} (ct_0/l)^{3/2} (\omega t_0)^{1/2} \sim 10^{22} kg/m^3$ at 1 Hz frequency).

One of the reasons for the status quo is that the argument presented above is based on sum-over-histories framework, which is not developed well in Causal Sets. The main area of development of "`quantum causet dynamics"' is the so-called \textit{sequential growth} model. [5] In the sequential growth model, universe is "`evolved"' from a single-element state by repeatedly adding elements. At $N$-th step, it can be described as a superposition of causal sets $C$ of cardinality $N$. There is a real-valued probability associated with each causal set (generalization to complex-valued amplitudes is in development). We go from $N$ to $N+1$ by enumerating all possible ways of adding one element to every causet and associating a transition probability with each addition.[6]


This is clearly not very different from the sum-over-histories approach. Both approaches give us structure of the "`Hilbert space"' of the universe in terms of relative likelihoods and phases of its possible states. We should expect that suitably defined sum over histories will either produce same results as SGM or converge with it in the continuum limit. 

In sum-over-histories formulation, $\Lambda$ is a variable that varies from history to history, impacting relative phases and amplitudes of histories. It's can't be a derivative of causal set structure, because inclusion of the term $\int \Lambda dV = \Lambda V$ into the action would mean that the theory is not even approximately local. Consequently, it is external: one could have two different histories with identical structures of causal relations, differing only in values of $\Lambda$ (and, consequently, with different amplitudes). Inclusion of $\Lambda$ into sequential growth model should be done on the same ground. Instead of viewing the N-element universe as a superposition of N-element causal sets, we should view it as a superposition of \textit{ordered pairs} $(C(N), \Lambda)$.  The amplitude to go from $(C(N), \Lambda)$ to $(C'(N+1), \Lambda')$ should depend on $\Lambda$ and $\Lambda'$. States with extreme deviations of cosmological constant from its expected value (likely zero) will somehow get suppressed during evolution. 

Consider an observer in an N-element universe. If he were to measure the cosmological constant, he could come up with any number that's consistent with background physics. Before the measurement is made, it only can be said that he is unlikely to observe $|\Lambda| \gg N^{-1/2}$. It is in this sense that the cosmological constant "`fluctuates"'. 
Once he has done the measurement, however, he is restricted to the subset of states of his universe that agree with his results. Different kinds of measurements may differ in their abilities to narrow down possible values of the cosmological constant, but they can't contradict each other (since distinct $\Lambda$'s live in orthogonal subspaces of the "`Hilbert space"', the probability to obtain two distinct values of $\Lambda$ in two different measurements is zero).
Consequently, the observer will be unable to detect any fluctuations in $\Lambda$ throughout his history. Once measured, the cosmological constant is really constant, from Big Bang to present time. 

There is one caveat to this argument. $\Lambda$ is not a quantity that can be measured directly, because it does not live inside a causet. It merely regulates which causets are probable and which are not. Therefore, precision of its measurement is limited by quantum fluctuations of spacetime. These fluctuations are, however, so tiny (no more than 1 part in $10^{15}$ at any time since electroweak phase transition in the early Universe) that they can be completely ignored for the purposes of this letter.



This result is somewhat unpleasant from experimental perspective, because it limits the possibility to confirm or falsify the theory of causal sets until its quantum dynamics is completely understood. From the theory standpoint, however, it is informative because it demonstrates that sequential growth dynamics should be modified in order to fully account for the presence of cosmological constant.

In closing, the following should be noted. It is common in discrete gravity theories to replace $dV$ with $dN$ when discretizing the path integral. This has a natural interpretation: elements are basic degrees of freedom, volume is an emergent feature and it does not belong in the elementary sum-over-histories formula. Replacing $\int \Lambda dV$ with $\int \Lambda dN$ completely negates the Causal Sets argument for $\Lambda$, which hinges on existence of fluctuations in the quantity $\Lambda$ is being multiplied with. We need to assume that 4-volume enters our sum over histories at the most fundamental level if we want to arrive at the desired result.
 
Yet it does \textit{not} mean that Causal Set theory is inconsistent unless the aesthetically unappealing action of "`forcing"' both variable $\Lambda$ and explicit 4-volume into the theory is taken. Cosmological constant of correct magnitude may still arise in the universe via some other mechanism; see [7], [8] for an interesting thermodynamical argument to this effect. 

\ReferencesBegin

[1] Dynamics of Dark Energy, Edmund J. Copeland, M. Sami, and Shinji Tsujikawa, hep-th/0603057

[2] R. D. Sorkin, in Relativity and Gravitation: Classical and Quantum, eds. J. C. D'Olivo et al. (World Scientific, Singapore, 1991); Int. J. Th. Phys. 36, 2759 (1997)

[3] "`Everpresent $\Lambda$"'; Maqbool Ahmed, Scott Dodelson, Patrick B. Greene, and Rafael Sorkin, astro-ph/0209274

[4] "`A Strong Constraint on Ever-Present Lambda"', John D. Barrow, gr-qc/0612128

[5] "`Causal Sets: Discrete Gravity"', Rafael D. Sorkin, gr-qc/0309009

[6] "`Relativity theory does not imply that the future already exists: a counterexample"', Rafael D. Sorkin, gr-qc/0703098

[7] "`Cosmological constant and vacuum energy"', G.E. Volovik, gr-qc/0405012

[8] "`The Universe in a Helium Droplet"', G.E. Volovik, Oxford University Press, ISBN 0198507828

\end{document}